\documentclass[conference]{IEEEtran}
\usepackage{ifpdf}

\IEEEoverridecommandlockouts
\usepackage{stfloats,color}
\usepackage{amsfonts}
\usepackage[cmex10]{amsmath}
\usepackage{graphicx}
\usepackage{setspace}
\usepackage{cite}
\usepackage{array}
\usepackage{algorithmic}
\usepackage{mdwmath}
\usepackage{mdwtab}
\usepackage{amssymb}
\usepackage{caption}
\usepackage{subcaption}
\usepackage[font=footnotesize]{subfig}
\usepackage{fixltx2e}
\usepackage{url}
\usepackage{times}
\usepackage{epsfig}
\usepackage{latexsym}
\usepackage{epstopdf}
\usepackage{verbatim}
\usepackage{units}
\usepackage{amsthm}
\usepackage{mdwlist}
\usepackage{lipsum,amsmath,multicol}
\usepackage{soul}
\graphicspath{{./figures/}}

\newtheorem{thm}{Theorem}
\newtheorem*{prooof}{Proof}
\newtheorem{lemm}{Lemma}

\begin{document}
%
\title{The Deterministic Capacity of Relay Networks with Relay Private Messages}



%
\author{\IEEEauthorblockN{Ahmed A. Zewail\IEEEauthorrefmark{1},
Y. Mohasseb\IEEEauthorrefmark{2},
M. Nafie\IEEEauthorrefmark{1} and 
H. EL Gamal \IEEEauthorrefmark{3} }
\IEEEauthorblockA{\IEEEauthorrefmark{1}Wireless Intelligent Networks Center (WINC), Nile University, Giza, Egypt \\ 
     Email:  ahmed.zewail@nileu.edu.eg,\ mnafie@nileuniversity.edu.eg}
\IEEEauthorblockA{\IEEEauthorrefmark{2}Department of Communications, The Military Technical College, Cairo, Egypt 11331\\
     Email: mohasseb@ieee.org}
\IEEEauthorblockA{\IEEEauthorrefmark{3}ECE Department, The Ohio State University, Columbus, OH\\
     Email: helgamal@ece.osu.edu}}


\maketitle

\begin{abstract}
 We study the capacity region of a deterministic 4-node network, where 3 nodes can only communicate via the fourth one. However, the fourth node is not merely a relay since it can exchange private messages with all other nodes. This situation resembles the case where a \textit{base station} relays messages between users and delivers messages between the backbone system and the users. We assume an asymmetric scenario where the channel between any two nodes is not reciprocal. First, an upper bound on the capacity region is obtained based on the notion of single sided genie. Subsequently, we construct an achievable scheme that achieves this upper bound using a superposition of broadcasting node 4 messages and an achievable "detour" scheme for a reduced 3-user relay network. 
\end{abstract}
\footnotetext[1]{This paper was made possible by NPRP grant $\#$ 4-1119-2-427 from the Qatar National Research Fund (a member of Qatar Foundation). The statement made herein are solely responsibility of the authors.}
\footnotetext[2]{Mohammed Nafie is also affiliated with the Department of Electronics and Communications, Cairo University.}



%
\IEEEpeerreviewmaketitle

\section{Introduction}

\IEEEPARstart{T}\small{h}e deterministic channel model for the wireless channel is used to approximate the capacity of Gaussian relay networks \cite{avestimehr2011wireless}. This model simplifies the wireless network interaction model by eliminating the noise, where the least significant bits are truncated at noise level. Thus, it allows us to focus only on the interaction between signals, and to envision schemes to maximize achievable rate tuples. Insights gleaned from these schemes can be used to find close approximations of the Gaussian capacity. In \cite{avestimehr2009capacity}, this model was applied to the multi-pair bidirectional relay network, which can be considered as a generalization of the bidirectional relay channel and its capacity region was analyzed and the cut set upper bound was shown to be tight. The authors also devised a simple equation-forwarding strategy that achieves this capacity region, in which different pairs are orthogonalized on the signal level space and the relay just re-orders the received equations created from the superposition of the transmitted signals on the wireless medium and forwards them. In \cite{hassibi2009approximate}, the authors studied the capacity of the Gaussian two-pair full-duplex bi-directional relay network. They proposed a transmission strategy which is based on a certain superposition of lattice codes and random Gaussian codes at the source nodes.\newline In \cite{mokhtar2010deterministic}, the authors studied the X-Channel, which represents a two user bidirectional half duplex wireless relay network. First, they developed a new upper bound based on the notion of a single-sided genie, then they used it to characterize the multicast capacity of their network. To prove the achievability, they proposed the idea of detour schemes that route some bits intended for a certain receiver via alternative paths when they cannot be accommodated on direct routes. Then, the capacity region of the deterministic Y-channel was found in \cite{chaaban2011capacity}. The authors established the achievability by using three schemes: bi-directional, cyclic, and uni-directional communication. In \cite{zewail2013deterministic}, we studied the capacity of the deterministic 4-user relay network with no direct links. We developed a new upper bound based on the notion of single sided genie, and proved its achievability via using one of two detour schemes that differ from the one in \cite{mokhtar2010deterministic} due to the different nature of our network.\newline
In this work, we characterize the capacity region of a deterministic 4-node network, where there is no direct links between 3 of them, thus they can only communicate via the forth one. In other words our network can be considered as 3-user relay network, where the relay is interested in exchanging some private messages with the three other users. Unlike the work done in [4-6], we consider the asymmetric case where the uplink and downlink channels are not reciprocal. To establish achievability, the scheme delivers the private messages of the fourth node  first, reducing the remaining task to sending the messages of the other 3 nodes via the "relay". This network thus becomes a 3-user asymmetric channels and reduced rates. Therefore, we characterize the capacity of this reduced network, then prove its achievability through a combination of a simple ordering scheme \cite{avestimehr2009capacity} and a special detour scheme.\\
Following this introduction, Section \ref{mainresult} outlines the main system assumptions and states the capacity region of our network. The first part of our new achievability scheme where we deal with the messages related to the fourth node is detailed in Section \ref{achiev}. To continue our achievability proof, we study the capacity of the asymmetric 3-user relay networks and its achievability  in Section \ref{3node}. Then, in Section \ref{onegeine}, we show the development of the upper bound based on the notion of single sided genie. Section \ref{ex} reports a numerical example to illustrate our achievability schemes. Finally, Section \ref{con} presents our conclusions.
\vspace{-4pt}
\section{The Main Result}\label{mainresult}
\vspace{-4pt}
Consider the network shown in Fig. \ref{fig:sm}. It consists of 4 nodes, where each node can exchange private messages with the three other nodes. Nodes 1 to 3 have no direct nodes between them and can only communicate via node 4. We assume that the channel between each node and the node 4 is not reciprocal. According to the deterministic channel model in \cite{avestimehr2011wireless}, this means that the number of levels between node $i$ and node 4, are not the same in the uplink and downlink i.e. ($ n_{i4} \neq n_{4i}$) where $n =\lceil 0.5 \log SNR \rceil$ and $i$ $\in$ $\{1,2,3\}$. \newline
 In the uplink phase, nodes $1, 2$ and $3$ transmit their messages to node $4$ which is only receiving in this phase, while in the downlink phase, node 4 acts as a relay for the private messages between the other three nodes and transmits its own messages to them.\newline
It follows from the assumption of non-reciprocal channels that the uplink and downlink channel gains do not have the same order. Therefore, we assume without loss of generality  $n_{u4}\geq$ $n_{v4}\geq$ $n_{t4}$ and $n_{4w}\geq$ $n_{4y}\geq$ $n_{4z}$, where $\{u,v,t\}, \{w,y,z\} \in \{1,2,3\}$.\newline
This paper derives the capacity region of this network, which is given by the following theorem:
\begin{figure}
\includegraphics[width=0.45\textwidth,height=0.13\textheight]{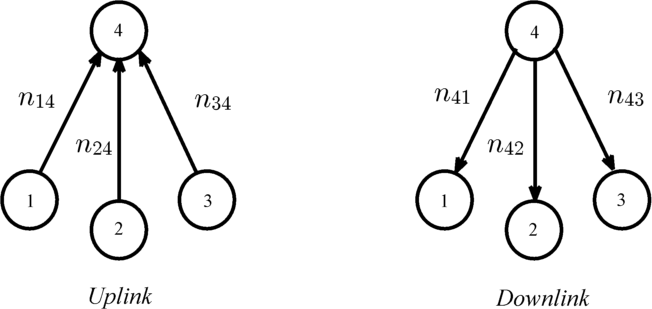}
\centering
\caption{System model}\label{fig:sm}
\end{figure}
\begin{thm}\label{upperbound}
The capacity region of our network is given by the rates that satisfy the following conditions:
\small
\begin{equation}\label{UL1}
R_{tu}+R_{tv}+R_{t4} \leq n_{t4}
\end{equation}
\begin{equation}\label{DL1}
R_{wz}+R_{yz}+R_{4z} \leq n_{4z}
\end{equation}
\begin{equation}\label{UL2}
R_{t4}+R_{v4}+R_{tu}+R_{vu}+\max(R_{tv},R_{vt})\leq n_{v4} 
\end{equation}
\begin{equation}\label{DL2}
R_{4z}+R_{4y}+R_{wz}+R_{wy}+\max(R_{zy},R_{yz})\leq n_{4y}
\end{equation}
\begin{equation}\label{UL3}
R_{u4}+R_{v4}+R_{t4}+R_{uv}+R_{ut}+\max(R_{tv},R_{vt})\leq n_{u4}
\end{equation}
\begin{equation}\label{UL4}
R_{u4}+R_{v4}+R_{t4}+R_{vu}+R_{vt}+\max(R_{tu},R_{ut})\leq n_{u4}
\end{equation}
\begin{equation}\label{UL5}
R_{u4}+R_{v4}+R_{t4}+R_{tu}+R_{tv}+\max(R_{uv},R_{vu})\leq n_{u4}
\end{equation}
\begin{equation}\label{DL3}
R_{4w}+R_{4y}+R_{4z}+R_{wz}+R_{yz}+\max(R_{wy},R_{yw})\leq n_{4w} 
\end{equation}
\begin{equation}\label{DL4}
R_{4w}+R_{4y}+R_{4z}+R_{yw}+R_{zw}+\max(R_{zy},R_{yz})\leq n_{4w} 
\end{equation}
\begin{equation}\label{DL5}
R_{4w}+R_{4y}+R_{4z}+R_{wy}+R_{zy}+\max(R_{wz},R_{zw})\leq n_{4w} 
\end{equation}
\normalsize
where $R_{ij}$ is the rate from node $i$ to node $j$.
\end{thm}
\section{Achievability}\label{achiev}
\begin{figure*}
\centering
\begin{subfigure}{.48\textwidth}
  \centering
  \includegraphics[width=1\linewidth]{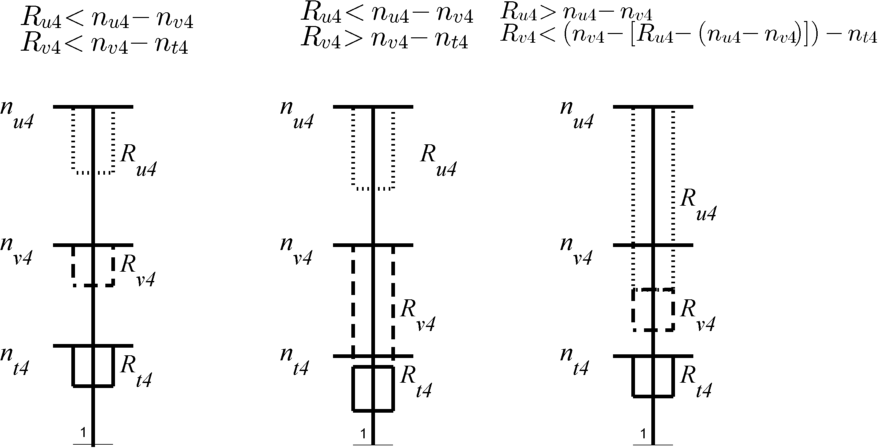}
  \caption{Uplink: different assignments are shown for the cases represented by the inequalities above the sub figures.}
  \label{fig:sub1}
\end{subfigure}%
\begin{subfigure}{.48\textwidth}
  \centering
  \includegraphics[width=.95\linewidth]{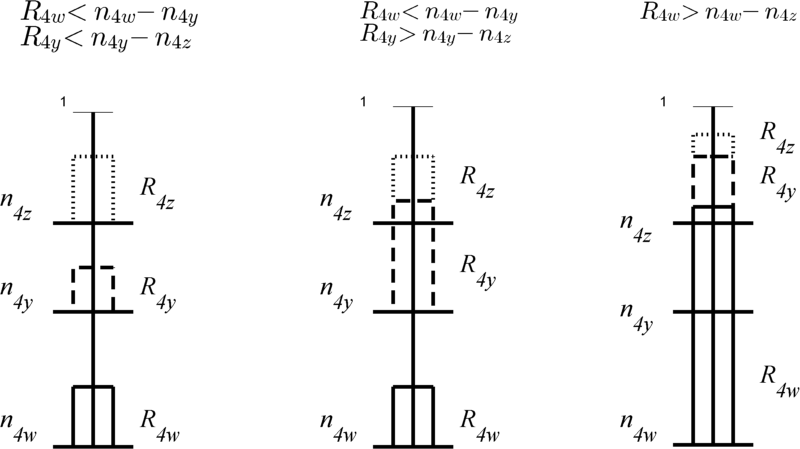}
  \caption{Downlink: different assignments are shown for the cases represented by the inequalities above the sub figures.}
  \label{fig:sub2}
\end{subfigure}
 \centering
\caption{Assigning levels for node 4 messages. The dotted, dashed and solid rectangles represent levels assigned for  $R_{u4}$, $R_{v4}$ and $R_{t4}$ in the uplink (a) and for $R_{4z}$, $R_{4y}$ and $R_{4w}$ in the downlink (b) respectively.}
\label{levels}
\end{figure*}

\vspace{-.07 in}
Our achievability scheme can be divided into two main parts: in the first part, we deal with the messages related to node 4, then we obtain a reduced network in the form of a 3-user relay network. Therefore, in the second part of our achievability scheme we deal with an asymmetric 3-user relay network, thus we need to generalize the work done for reciprocal Y-channel in \cite{chaaban2011capacity} by considering an asymmetric scenario where $n_{iR} \neq n_{Ri}$.\newline 
In the uplink phase, we assign the highest $R_{i4}$ levels from $n_{i4}$ levels to the messages from node $i$ to node $4$, while in the downlink phase we assign the lowest $R_{4i}$ levels from $n_{4i}$ levels to the messages from node $4$ to node $i$. \newline
Note that if $R_{u4}> n_{u4}-n_{v4}$, then some of the assigned levels to $R_{u4}$ can interfere with the levels carrying bits from node $v$ to node 4. Therefore, for $R_{v4}$, the assignment begins from the highest remaining level from node $v$ to the node 4 after the assignment of the levels required to serve $R_{u4}$. This case and other cases are shown in Fig. \ref{levels}.\newline
The above operation of assigning the levels for the messages related to node 4 can be represented by subtracting rates associated with node 4 from both sides of the inequalities in Theorem 1. Let $n_{uR}=n_{u4}-R_{u4}-R_{v4}-R_{t4}$ and $n_{Rw}=n_{4w}-R_{4w}-R_{4y}-R_{4z}$.\newline 
After subtracting these rates from (\ref{UL2}) and (\ref{UL4}), we have the following conditions: 
\begin{equation}\nonumber
R_{tu}+R_{vu}+R_{vt}\leq n_{v4}-R_{t4}-R_{v4} 
\end{equation}
\begin{equation}\nonumber
R_{tu}+R_{vu}+R_{vt}\leq n_{u4}-R_{u4}-R_{v4}-R_{t4}
\end{equation}
Now, these two conditions can be combined as 
\begin{equation}\label{red1}
R_{vu}+R_{vt}+R_{tu}\leq n_{vR}
\end{equation}
where 
\begin{equation}\nonumber
\begin{aligned}
 n_{vR}&=\min\{n_{v4}-R_{t4}-R_{v4},n_{u4}-R_{u4}-R_{v4}-R_{t4}\} \\ 
 &=n_{v4}-R_{t4}-R_{v4}-[R_{u4}-(n_{u4}-n_{v4})]^+
\end{aligned}
\end{equation}
Again, by applying the same process on (\ref{UL2}) and (\ref{UL5}), we get 
\begin{equation}\label{red2}
R_{vu}+R_{tv}+R_{tu}\leq n_{vR}
\end{equation} 
We can combine the conditions (\ref{red1}) and (\ref{red2}) to get (\ref{vR}).\newline 
Also, by subtracting rates related to node 4 from (\ref{UL2}) and (\ref{UL5}), we get 
\begin{equation}\nonumber
R_{tu}+R_{tv}+R_{vu}\leq n_{v4}-R_{t4}-R_{v4} 
\end{equation}
\begin{equation}\nonumber
R_{tu}+R_{tv}+R_{vu}\leq n_{u4}-R_{u4}-R_{v4}-R_{t4}
\end{equation}
Since $R_{vu}\geq 0$, then this implies 
\begin{equation}\label{tR1}
R_{tu}+R_{tv}\leq n_{v4}-R_{t4}-R_{v4} 
\end{equation}
\begin{equation}\label{tR2}
R_{tu}+R_{tv}\leq n_{u4}-R_{u4}-R_{v4}-R_{t4} \triangleq n_{uR}
\end{equation}
And by subtracting rates related to node 4 from (\ref{UL1}), we get
\begin{equation}\label{tR3}
R_{tu}+R_{tv}\leq n_{t4}-R_{t4}
\end{equation} 
Again, the conditions (\ref{tR1})-(\ref{tR3}) can be combined as
\begin{equation}\nonumber
R_{tu}+R_{tv}\leq n_{tR}
\end{equation} 
which is condition (\ref{tR}), where
\begin{equation}\nonumber
\begin{aligned}
 n_{tR}&=\min\{(n_{t4}-R_{t4}),(n_{v4}-R_{t4}-R_{v4}),n_{uR}\} \\ 
 &=n_{t4}-R_{t4}-\beta
\end{aligned}
\end{equation}
\begin{figure*}[ht]
\[\beta =\begin{cases} [R_{u4}-(n_{u4}-n_{v4})]^+ +R_{v4}  &  \mbox{ $ R_{u4}\geq (n_{u4}-n_{t4})$} \\
 [R_{v4}-(n_{v4}-n_{t4}-[R_{u4}-(n_{u4}-n_{v4})]^+)]^+ &  \mbox{ $ R_{u4}\geq (n_{u4}-n_{v4})$ and $R_{u4}<(n_{u4}-n_{t4})$}\\
 [R_{v4}-(n_{v4}-n_{t4})]^+ &  \mbox{ $ R_{u4}< (n_{u4}-n_{v4})$}
  \end{cases}
\]
\[\gamma =\begin{cases} [R_{4w}-(n_{4w}-n_{4z})]^+ +R_{4y} &  \mbox{ $ R_{4w}\geq (n_{4w}-n_{4z})$} \\
 [R_{4y}-(n_{4y}-n_{4z}-[R_{4w}-(n_{4w}-n_{4y})]^+)]^+ &  \mbox{ $ R_{4w}\geq (n_{4w}-n_{4y})$ and $R_{4w}<(n_{4w}-n_{4z})$}\\
 [R_{4y}-(n_{4y}-n_{4z})]^+ &  \mbox{ $ R_{4w}< (n_{4w}-n_{4y})$}
  \end{cases}
 \]
\end{figure*}
Also, by applying this subtraction on the conditions (\ref{UL3})-(\ref{DL5}), we obtain the conditions (\ref{start})-(\ref{end}). \newline
Proceeding similarly for the remaining downlink levels we obtain the conditions (\ref{Ry}) and (\ref{Rz}), where  
\begin{equation}\nonumber
n_{Ry}=n_{4y}-R_{4z}-R_{4y}-[R_{4w}-(n_{4w}-n_{4y})]^+
\end{equation}
\begin{equation}\nonumber
n_{Rz}=n_{4z}-R_{4z}-\gamma
\end{equation}
From the above expressions of the reduced channel gains, it is clear that $n_{uR}\geq n_{vR}\geq n_{tR}$ and $n_{Rw}\geq n_{Ry}\geq n_{Rz}$.\newline
After serving messages related to node $4$, we can put $R_{u4}=R_{v4}=R_{t4}=0$ and $R_{4w}=R_{4y}=R_{4z}=0$. Now, we have a reduced capacity region which is defined by the conditions stated in Theorem 2 in the following section, we can observe that this region is equivalent to the capacity region of a 3-user relay network with channel gains  $n_{uR}, n_{vR}$ and $n_{tR}$ in the uplink and  $n_{Rw}, n_{Ry}$ and $n_{Rz}$ in the downlink. Therefore, to continue our achievability proof for our original network, we need to prove the achievability of the reduced region stated in Theorem 2 in the next section.
\section{Asymmetric 3-user relay network}\label{3node}
As we mentioned before, the reduced capacity region after serving the  messages related to node 4 is equivalent to the one of an asymmetric 3-user relay network. Therefore, we need to generalize the capacity region in \cite{chaaban2011capacity} and prove its achievability. The capacity region of this reduced network is defined by the following theorem:
\begin{thm}\label{3upperbound}
The capacity region of the deterministic 3-user asymmetric relay network is given by the rates that satisfy the following inequalities:
\small
\begin{equation}\label{tR}
R_{tu}+R_{tv} \leq n_{tR}
\end{equation}
\begin{equation}\label{Rz}
R_{wz}+R_{yz} \leq n_{Rz}
\end{equation}
\begin{equation}\label{vR}
R_{tu}+R_{vu}+\max(R_{tv},R_{vt}) \leq n_{vR}
\end{equation}
\begin{equation}\label{Ry}
R_{wz}+R_{wy}+\max(R_{yz},R_{zy}) \leq n_{Ry}
\end{equation}
\begin{equation}\label{start}
R_{uv}+R_{ut}+\max(R_{tv},R_{vt})\leq n_{uR}
\end{equation}
\begin{equation}
R_{vu}+R_{vt}+\max(R_{tu},R_{ut})\leq n_{uR}
\end{equation}
\begin{equation}
R_{tu}+R_{tv}+\max(R_{uv},R_{vu})\leq n_{uR}
\end{equation}
\begin{equation}
R_{wz}+R_{yz}+\max(R_{wy},R_{yw})\leq n_{Rw}
\end{equation}
\begin{equation}
R_{yw}+R_{zw}+\max(R_{zy},R_{yz})\leq n_{Rw} 
\end{equation}
\begin{equation}\label{end}
R_{wy}+R_{zy}+\max(R_{wz},R_{zw})\leq n_{Rw} 
\end{equation}
\end{thm}
\normalsize
\subsection{Achievability of asymmetric 3-user relay network}\label{Scheme}
The achievability of any rate tuple that satisfies the inequalities in Theorem 2 can be attained using one of two schemes: the Simple Ordering Scheme (SOS) or a Detour Scheme (DS) which converts our network to an equivalent one with modified rates to which we can apply the SOS. It should be mentioned that our detour scheme is different from that in \cite{mokhtar2010deterministic} due to the different nature of the problem. Therefore, in the following subsections, we will focus on this new detour scheme and only highlight the SOS.
\subsubsection {The Simple Ordering Scheme (SOS)}
In [2 and 4] the bits from different sources are placed so that they combine at the relay in such a way to allow decoding at the recipients. In the uplink, the nodes align their bits so that bits transmitted from node $i$ to node $j$ i.e. ($x_{ij}$) are xored with bits transmitted from $j$ to $i$ i.e. ($x_{ji}$) at the relay. In the downlink phase, since the relay does not need to decode each bit individually, it can simply broadcast $x_{ij} \oplus x_{ji}$, and since node $i$ knows $x_{ij}$, it can decode $x_{ji}$, and vice verse for node $j$.
\begin{lemm}\label{SOS3}
The Simple Ordering Scheme (SOS) can achieve all the integral rate tuples that simultaneously satisfy the conditions stated in Theorem \ref{3upperbound} and the following two conditions:
\begin{equation}\label{eqndetour31}
R_{12}+R_{23}+R_{31}\leq \min(n_{uR},n_{Rw})
\end{equation}
\begin{equation}\label{eqndetour32}
R_{21}+R_{13}+R_{32}\leq \min(n_{uR},n_{Rw})
\end{equation}
\end{lemm}
\hspace{-.25 in}
\textbf{Proof Sketch.} SOS can only work if node $i$ is able to transmit its data on the available number of levels $n_{iR}$ and receive its data on the available number of levels $n_{Ri}$. Our proof depends on finding the size of each of the three segments corresponding to the messages from each node, in a similar way that we followed in the proof of Lemma 1 in subsection III-A in \cite{zewail2013deterministic}. Subsequently, by comparing the resultant conditions in both uplink and downlink phases with the conditions in Theorem \ref{3upperbound}, we get the extra conditions needed to apply the SOS as stated in Lemma 1.
\subsubsection {The Detour Scheme (DS)}\label{detourScheme}
If any achievable rate tuple that satisfies the conditions stated in Theorem \ref{3upperbound}, violates one of the two extra conditions stated Lemma \ref{SOS3}, the SOS will not work. In this case, we use our detour scheme to convert the network into an equivalent one, where the conditions in Theorem \ref{3upperbound} and Lemma \ref{SOS3} are satisfied simultaneously, thus we can apply the SOS on this equivalent network.\newline 
It is clear that each one of the two conditions stated in Lemma \ref{SOS3} consists of a 3-node cycle which is represented by the data flow of the rates in the left hand side (LHS). Thus, these two extra conditions represents the cycle between the three nodes in the two possible directions.
To explain this scheme, we assume without loss of generality that for a certain $\{i,j,k\} \in \{1,2,3\}$.
\begin{equation}\nonumber
R_{ij}+R_{jk}+R_{ki}> \min(n_{uR},n_{Rw})
\end{equation}
Now, we need to reduce the rates of the left hand side (LHS) by subtracting $\lambda$ bits, such that the rates modified rates satisfy:
\begin{equation}\nonumber
(R_{ij}+R_{jk}+R_{ki})-\lambda \leq \min(n_{uR},n_{Rw})
\end{equation}
The subtracted $\lambda$-bits will be transmitted to their respective destination via alternate paths. Thus, all rates along this detour must be increased, while at the same time satisfying the other conditions in Theorem \ref{3upperbound} and Lemma \ref{SOS3}. For example, if we decide to detour $\lambda$-bits from the $R_{ij}$ via node $k$, this follows that each rate of $R_{ik}$ and $R_{kj}$ should be increased by $\lambda$. Therefore, whichever the rate we choose to detour this $\lambda$-bits from, the rates over the reverse cycle should be modified as:
\vspace{-.05 in} 
\begin{equation}\nonumber
R_{ji}+R_{ik}+R_{kj}\rightarrow R_{ji}+R_{ik}+R_{kj}+2\lambda
\end{equation} 
\begin{lemm}
For all integer rate tuples for the 3-user relay network satisfying Theorem \ref{3upperbound} and violating only one of the two conditions in Lemma \ref{SOS3}, it is possible to modify the rates using the detour scheme and find an equivalent network, that can achieve the original rate tuple via alternative paths.
\end{lemm}
\begin{prooof}
See the appendix.  
\end{prooof}
\begin{lemm}
There are no achievable rate tuples that violate the two conditions (\ref{eqndetour31}) and (\ref{eqndetour32}) simultaneously, which represent a cycle in its two possible directions.
\end{lemm}
\hspace{-.25 in}
\textbf{Proof Sketch. }Assuming that the two conditions in Lemma \ref{SOS3} are violated simultaneously, by summing these two conditions and multiplying the resultant condition by 3, we obtain
\begin{equation}\nonumber
3R_{12}+3R_{21}+3R_{13}+3R_{31}+3R_{23}+3R_{32}>6\min(n_{uR},n_{Rw})
\end{equation}
However, from combining the conditions (\ref{start})-(\ref{end}), we can get three conditions restricted by $\min(n_{uR},n_{Rw})$. By summing them and multiplying the resultant condition by 2, we get 
\vspace{-0.07 in}
\begin{multline}\nonumber
2R_{12}+2R_{21}+2R_{13}+2R_{31}+2R_{23}+2R_{32}+2\max(R_{12},R_{21})\\+2\max(R_{13},R_{31})+2\max(R_{23},R_{32})\leq 6\min(n_{uR},n_{Rw})
\end{multline}
It is straightforward to prove that the LHS of this inequality is larger than the previous one, resulting in a contradiction. Thus, it is clear that if the two extra conditions in Lemma \ref{SOS3} are violated simultaneously, the upper bound will be also violated.
\section{An upper bound Based on Single sided genie}\label{onegeine}
Classical cut set bounds divide the nodes of a network two sets  $S$ and $S^c$ which represent the transmitting and receiving nodes respectively \cite{cover2006elements}. Nodes in each of these sets are assumed to fully cooperate by sharing their side information. Hence, such bounds will be referred to as the two sided genie aided bounds. In \cite{mokhtar2010deterministic}, it was shown that applying these traditional cut set bounds to the relay network produces loose results. A tighter single sided genie bound was developed, which will not be iterated here. Instead, we will focus only on some points arising from the different nature of the network under consideration.
\subsection{For the 4-Node Network}
The single sided genie upper bound can be found by considering all possible cuts through the network, then all the orders in which the single sided genie works. Then combining the resulting inequalities to minimize the set of conditions. A simplifying observations, however, is that we need not take into account all cuts where $S$ contains node 4 with any other node, which significantly reduces the number of inequalities to consider. For example, let $S=\{4,1\}$ and let the genie transfers only the data from node $1$ to node $4$ i.e. ($R_{14}$), this will provide the following bound
\vspace{-.1 in} 
\begin{equation}\nonumber
R_{42}+R_{43}+R_{12}+R_{13}+R_{41}\leq \max(n_{41},n_{42},n_{43})+n_{14} 
\end{equation}
This condition is implicitly satisfied from the resultant bounds from the cuts $S=\{4\}$ and $S=\{1\}$, therefore this cut does not result in a new constraint on the capacity region.
\subsection{For the 3-User Asymmetric Relay Network}
For the two sided genie bound, the cut around the relay i.e. ($S=\{R\}$) is not useful, as the relay does not have its own rates to transmit. In contrast, the one sided genie can be calculated for different genie orders. For example in the uplink phase, if we assume the genie order $i \rightarrow j \rightarrow k $, which means that the genie transfers all data of node $i$ to nodes $j$ and $k$, then the data of node $j$ to nodes $k$, therefore the one sided genie bound in this case will be as follows:
\vspace{-0.07 in} 
\begin{equation}\nonumber
R_{ki}+R_{kj}+R_{ji}\leq \max(n_{1R},n_{2R},n_{3R})
\end{equation} 
\vspace{-0.25 in}
\section{Numerical Example}\label{ex}
\begin{figure}
\includegraphics[width=0.4\textwidth,height=0.15\textheight]{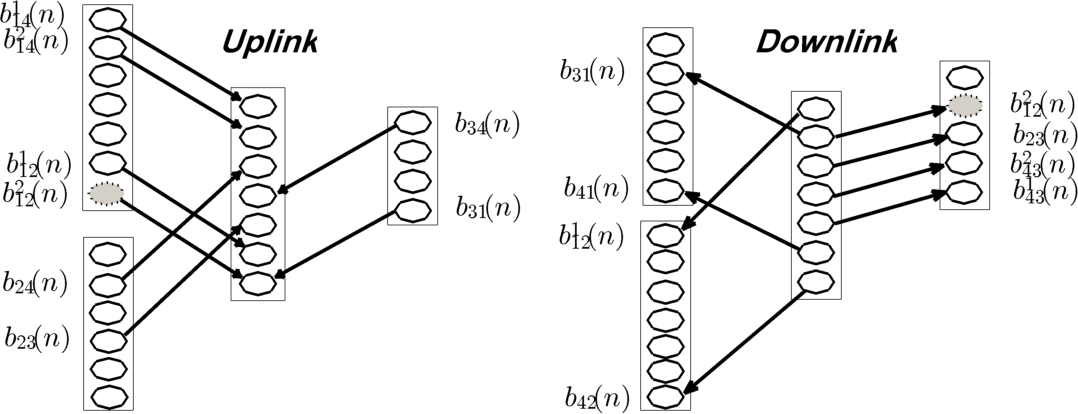}
\centering
\includegraphics[width=0.40\textwidth,height=0.15\textheight]{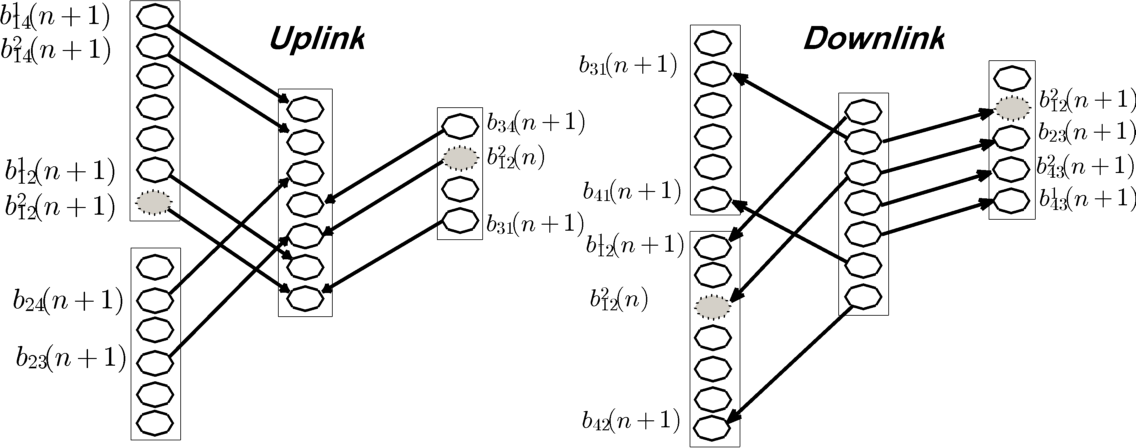}
\centering
\caption{An example that illustrates our achievability scheme.}\label{exxxfig}
\end{figure}

Consider a network with channel gains $(n_{14},n_{24},n_{34})=(7,6,4)$ and $(n_{41},n_{42},n_{43})=(6,7,5)$ and a rate tuple $R=$ $(R_{12},R_{13},R_{14},R_{21},R_{23},R_{24},R_{31},R_{32},R_{34},R_{41},R_{42},R_{43})$ $=(2,0,2,0,2,1,1,0,1,1,1,1)$.\\
To show the achievability of this rate tuple, we first dedicate resources to the messages related to node 4.\newline In the uplink phase: following the procedure explained in Section \ref{achiev}, we assign the highest 2 levels of $n_{14}$ for $R_{14}$,  the second highest level of $n_{24}$ for $R_{24}$ and the highest level of $n_{34}$ for $R_{34}$, which follows the third case in Fig. \ref{fig:sub1}. Conversely, in the downlink phase: we assign the lowest level from $n_{42}$, $n_{41}$ and $n_{43}$ for $R_{42}$, $R_{41}$ and $R_{43}$ respectively, which follows the first case in Fig. \ref{fig:sub2}.\newline
Now, we have a reduced 3-user relay network with channel gains $(n_{1R},n_{2R},n_{3R})$ = (3,3,3) and $(n_{R1},n_{R2},n_{R3})$ = (4,4,4) and the rate tuple $R$ = $(R_{12},R_{13},R_{21},R_{23},R_{31},R_{32})$ = (2,0,0,1,1,0), which violates the extra condition $R_{12}+R_{23}+R_{31}$ = 3+1, therefore we will detour one bit from $R_{12}$ via node $3$, thus  
$R_{12}\rightarrow R_{12}-1$, $R_{13}\rightarrow R_{13}+1$ and $R_{32}\rightarrow R_{32}+1$ and the modified rate tuple is $\grave{R}=(1,1,0,1,1,1)$, where we can apply the SOS as shown in Fig. \ref{exxxfig} because this modified rate tuple satisfies the two extra SOS conditions.
\vspace{-0.1 in}
\section{Conclusions}\label{con}
We characterized of the multicast capacity region of a deterministic 3-user relay network where the relay is interested in exchanging some private messages with the network users. In the course of our achievability proof, we achieved the capacity of the asymmetric 3-user relay network. The upper bound was obtained using a single sided genie, and the capacity region was achieved using either a simple reordering scheme or a detour scheme. Based on our findings, we conjecture that if the capacity of the $K$-user relay network can be found, for any $K$, it follows that we can also find the capacity of the $K$-user relay network where where the relay is interested in exchanging some private messages with the network users by the same procedure we followed in this paper.
\section*{Appendix }
Let $n^*$=$\min(n_{uR},n_{Rw})$, and assume that the condition in (\ref{eqndetour31}) is violated by $\lambda$ bits, then we have 
\begin{equation}\label{violated}
R_{12}+R_{23}+R_{31}=n^*+\lambda
\end{equation}
However, from the conditions (\ref{start})-(\ref{end}) in Theorem \ref{3upperbound}, we can get
\begin{equation}\label{apped} 
\begin{aligned}
&R_{12}+R_{13}+R_{23}\leq n^* \\ 
&  R_{21}+R_{23}+R_{31}\leq n^* \\
&  R_{31}+R_{32}+R_{12}\leq n^*
  \end{aligned}
\end{equation}
By comparing (\ref{apped}) with (\ref{violated}), we get 
\begin{equation}\label{eqn: R1icondUB}
\begin{aligned}
R_{31} \geq R_{13}+\lambda
& & R_{12} \geq R_{21}+\lambda
& & R_{23} \geq R_{32}+\lambda
\end{aligned}
\end{equation}
The choice of the rate from which we will subtract the $\lambda$ bits, depends on the order of the channel gains:
\subsubsection*{1) Detour bits from $R_{12}$}
We can detour bits from $R_{12}$ if  $n_{3R}$ and $n_{R3}$ are not the lowest channel gains in UL and DL respectively, and
the detour scheme will be
\begin{equation}\nonumber
\begin{aligned}
R_{12} \rightarrow R_{12}-\lambda
& & R_{13} \rightarrow R_{13}+\lambda
& & R_{32} \rightarrow R_{32}+\lambda
\end{aligned}
\end{equation}
\subsubsection*{2) Detour bits from $R_{23}$}
We can detour bits from $R_{23}$ if  $n_{1R}$ and $n_{R1}$ are not the lowest channel gains in UL and DL respectively, and the detour scheme will be
\begin{equation}\nonumber
\begin{aligned}
R_{23} \rightarrow R_{23}-\lambda
& & R_{21} \rightarrow R_{21}+\lambda
& & R_{13} \rightarrow R_{13}+\lambda
\end{aligned}
\end{equation}
\subsubsection*{3) Detour bits from $R_{31}$}
We can detour bits from $R_{31}$ if  $n_{2R}$ and $n_{R2}$ are not the lowest channel gains in UL and DL respectively, and the detour scheme will be
\begin{equation}\nonumber
\begin{aligned}
R_{31} \rightarrow R_{31}-\lambda
& & R_{32} \rightarrow R_{32}+\lambda
& & R_{21} \rightarrow R_{21}+\lambda
\end{aligned}
\end{equation}
From (\ref{eqn: R1icondUB}), and by checking the the conditions in Theorem \ref{3upperbound} and in Lemma \ref{SOS3} for the modified rate tuple, it can be shown from these conditions are satisfied.\newline
Therefore, regardless of the order of the channel gains in both uplink and downlink phases, at least one detour from the above three detours can be applied to convert the network into an equivalent one that satisfies the conditions in Theorem \ref{3upperbound} the extra SOS conditions.
\bibliographystyle{IEEEtran}
\vspace{-.1 in}
\bibliography{IEEEabrv,DiversityLib}

\begin{thebibliography}{1}
\providecommand{\url}[1]{#1}
\csname url@samestyle\endcsname
\providecommand{\newblock}{\relax}
\providecommand{\bibinfo}[2]{#2}
\providecommand{\BIBentrySTDinterwordspacing}{\spaceskip=0pt\relax}
\providecommand{\BIBentryALTinterwordstretchfactor}{4}
\providecommand{\BIBentryALTinterwordspacing}{\spaceskip=\fontdimen2\font plus
\BIBentryALTinterwordstretchfactor\fontdimen3\font minus
  \fontdimen4\font\relax}
\providecommand{\BIBforeignlanguage}[2]{{%
\expandafter\ifx\csname l@#1\endcsname\relax
\typeout{** WARNING: IEEEtran.bst: No hyphenation pattern has been}%
\typeout{** loaded for the language `#1'. Using the pattern for}%
\typeout{** the default language instead.}%
\else
\language=\csname l@#1\endcsname
\fi
#2}}
\providecommand{\BIBdecl}{\relax}
\BIBdecl

\bibitem{avestimehr2011wireless}
A.~Avestimehr, S.~Diggavi, and D.~Tse, ``Wireless {N}etwork {I}nformation
  {F}low: {A} {D}eterministic {A}pproach,'' \emph{Information Theory, IEEE
  Transactions on}, vol.~57, no.~4, pp. 1872--1905, 2011.

\bibitem{avestimehr2009capacity}
A.~Avestimehr, M.~Khajehnejad, A.~Sezgin, and B.~Hassibi, ``Capacity {R}egion
  of {T}he {D}eterministic {M}ulti-pair {B}i-directional {R}elay {N}etwork,''
  in \emph{Networking and Information Theory, ITW 2009. IEEE Information Theory
  Workshop on}.\hskip 1em plus 0.5em minus 0.4em\relax IEEE, 2009, pp. 57--61.

\bibitem{hassibi2009approximate}
B.~Hassibi, A.~Sezgin, M.~Khajehnejad, and A.~Avestimehr, ``Approximate
  {C}apacity {R}egion of {T}he {T}wo-pair {B}idirectional {G}aussian {R}elay
  {N}etwork,'' in \emph{Information Theory, ISIT 2009. IEEE International
  Symposium on}.\hskip 1em plus 0.5em minus 0.4em\relax IEEE, 2009, pp.
  2018--2022.

\bibitem{mokhtar2010deterministic}
M.~Mokhtar, Y.~Mohasseb, M.~Nafie, and H.~El~Gamal, ``On {T}he {D}eterministic
  {M}ulticast {C}apacity of {B}idirectional {R}elay {N}etworks,'' in
  \emph{Information Theory Workshop, ITW 2010}.\hskip 1em plus 0.5em minus
  0.4em\relax IEEE, 2010, pp. 1--5.

\bibitem{chaaban2011capacity}
A.~Chaaban and A.~Sezgin, ``The {C}apacity {R}egion of {T}he {L}inear {S}hift
  {D}eterministic {Y}-{C}hannel,'' in \emph{Information Theory Proceedings,
  ISIT 2011 IEEE International Symposium on}.\hskip 1em plus 0.5em minus
  0.4em\relax IEEE, 2011, pp. 2457--2461.

\bibitem{zewail2013deterministic}
A.~A. Zewail, Y.~Mohasseb, M.~Nafie, and H.~El~Gamal, ``The {D}eterministic
  {M}ulticast {C}apacity of 4-{N}ode {R}elay {N}etworks,'' {T}o appear in {\it
  IEEE International Symposium on Information Theory Proceedings (ISIT)}, 2013
  arXiv:1304.4464v1.

\bibitem{cover2006elements}
T.~Cover and J.~Thomas, \emph{Elements of information theory}.\hskip 1em plus
  0.5em minus 0.4em\relax John Wiley \& Sons, 2006.

\end{thebibliography}
\vspace{-.1 in}
\end{document}